\documentclass[a4paper,fleqn,usenatbib]{mnras}

\usepackage[T1]{fontenc}
\usepackage{ae,aecompl}

\usepackage{graphicx}	
\usepackage{amsmath}	
\usepackage{amssymb}	

\title[Compact H$\alpha$ emitting regions from CHARA]{The compact H$\alpha$ emitting regions of the Herbig Ae/Be stars HD 179218 and HD 141569 from CHARA spectro-interferometry}

\author[I. Mendigut\'\i{}a et al.]
{\parbox{\textwidth}{I. Mendigut\'\i{}a$^{1}$\thanks{E-mail: \texttt{I.Mendigutia@leeds.ac.uk}},
R.D. Oudmaijer$^{1}$,
D. Mourard$^{2}$ and
J. Muzerolle$^{3}$
\vspace{0.4cm}}
\\
\parbox{\textwidth}{
$^{1}$School of Physics and Astronomy, University of Leeds, Woodhouse Lane, Leeds, LS2 9JT, UK.\\
$^{2}$Universit\'{e} C\^{o}te d\'{}Azur, OCA, CNRS, Lagrange, Parc Valrose, B\^{a}t. H. FIZEAU, 06108 Nice cedex 02, France\\
$^{3}$Space Telescope Science Institute, 3700 San Martin Dr., Baltimore, MD, 21218, USA\\
}}

\date{Accepted 2016 September 28; Received 2016 September 28; in original form 2016 August 11}

\pubyear{2015}

\begin{document}
\label{firstpage}
\pagerange{\pageref{firstpage}--\pageref{lastpage}}
\maketitle

\begin{abstract}
This work presents CHARA/VEGA H$\alpha$ spectro-interferometry (R $\sim$ 6000, and $\lambda$/2B $\sim$ 1 $mas$) of HD 179218 and HD 141569, doubling the sample of Herbig Ae/Be (HAeBe) stars for which this type of observations is available so far. The observed H$\alpha$ emission is spatially unresolved, indicating that the size of the H$\alpha$ emitting region is smaller than $\sim$ 0.21 and 0.12 $au$ for HD 179218 and HD 141529 ($\sim$ 15 and 16 R$_*$, respectively). This is smaller than for the two other HAeBes previously observed with the same instrumentation. Two different scenarios have been explored in order to explain the compact line emitting regions. A hot, several thousand $K$, blackbody disc is consistent with the observations of HD 179218 and HD 141569. Magnetospheric accretion (MA) is able to reproduce the bulk of the H$\alpha$ emission shown by HD 179218, confirming previous estimates from MA shock modelling with a mass accretion rate of 10$^{-8}$ M$_{\odot}$ yr$^{-1}$, and an inclination to the line of sight between 30 and 50$\degr$. The H$\alpha$ profile of HD 141569 cannot be fitted from MA due to the high rotational velocity of this object. Putting the CHARA sample together, a variety of scenarios is required to explain the H$\alpha$ emission in HAeBe stars -compact or extended, discs, accretion, and winds-, in agreement with previous Br$\gamma$ spectro-interferometric observations.    
\end{abstract}




\section{Introduction}
\label{Sect:Intro}
Pre-main sequence (PMS) objects are optically visible young stars mostly surrounded by proto-planetary discs. Low- and intermediate-mass PMS stars (Classical T Tauri; CTT, and Herbig Ae/Be stars; HAeBe, respectively) show IR excess from circumstellar discs and were primarily defined from the presence of the H$\alpha$ line in emission, which is mainly associated with accretion and winds \citep{Muzerolle98,Kurosawa06,Lima10,Tambo14}.    

Despite the presence of the H$\alpha$ emission line is a defining characteristic of PMS stars, spatially resolved observations of this feature are very scarce. The bulk of the hydrogen recombination line is supposed to come from the inner parts of the disc at a few  stellar radii from the central star, for which spectro-interferometric observations become crucial to investigate this. However, most of such studies have focused on the near-IR Br$\gamma$ transition, mainly due to the wider access to near-IR interferometers. Br$\gamma$ spectro-interferometry of an increasing number of PMS stars \citep[around 20 bright and close HAeBes up to date, see e.g.][]{Kraus15} reveals that the emission comes from a few stellar radii, generally tracing magnetospherically driven accretion, outflows and/or Keplerian discs \cite[see e.g.][]{Kraus08,Eisner10,Mendi15b,Kurosawa16}. Concerning H$\alpha$, the only spectro-interferometric observations in PMS stars were those obtained with the CHARA/VEGA interferometer \citep{Mourard09} for AB Aur \citep{Rousselet10} and MWC 361 \citep{Benisty13}. The results found for both HAeBe stars are consistent with a spatially resolved non-spherical wind. In a recently accepted paper, \citet{Perraut16} revisit the problem on the H$\alpha$ emission in AB Aur from CHARA data with improved sensitivity and spatial resolution. The new results confirm the disc-wind origin, although concluding that a magnetospheric accretion component cannot be completely neglected.

In this work we present CHARA/VEGA spectro-interferometric observations in H$\alpha$ for two additional HAeBes, HD 179218 and HD 141569, doubling the sample so far available with this type of observations. Since we find that the H$\alpha$ emission is unresolved in these objects, physical scenarios other than extended winds are tested. Section \ref{Sect:Sample} includes some properties of the target stars, Sect. \ref{Sect:Observations} describes the observations and data reduction, Sect. \ref{Sect:Analysis} presents an analysis in terms of upper limits for the size of the line emitting regions, blackbody discs, and magnetospheric accretion, and Sect. \ref{Sect:Conclusions} summarizes the main conclusions.

\section{Target stars}
\label{Sect:Sample} 

Table \ref{Table:prop} shows the distance, stellar temperature, mass, radius, projected rotational velocity and inclination to the line of sight (90$\degr$ for an edge-on and 0$\degr$ for a pole-on disc) of HD 179218 and HD 141569. Both objects are young, bright \citep[$<$ 10 Myr and $R$-magnitudes of 7.00 and 7.25, respectively; see][and references therein]{Mendi11b} and they are relatively closeby HAe (A0) stars with similar stellar properties, although HD 141569 is a fast rotator. Whereas the disc inclination for HD 141569 was inferred from spatially resolved observations, the one for HD 179218 has a larger degree of uncertainty, which could range from 0$\degr$ to 45$\degr$  \citep{liu07,vanderplas08}. The corresponding value indicated in Table \ref{Table:prop} comes from a simple model fit to the CO(J=3-2) line profile.

\begin{table}
\centering
\renewcommand\tabcolsep{1.7pt}
\caption{Properties of the stars}
\label{Table:prop}
\begin{tabular}{lrrrrrr}
\hline\hline
Star& d & T$_*$ & M$_*$ & R$_*$ & $v sin i$ & $i$ \\
... &[pc]& [K]  &[M$_{\odot}$]&[R$_{\odot}$] &[km s$^{-1}$] & [$\degr$]\\ 
\hline
HD 179218 & 201$\pm$24 & 9500$\pm$500 & 2.6$\pm$0.2 & 2.9$\pm$0.9 & 72$\pm$3 & 40$\pm$10\\
HD 141569 & 112$\pm$34 & 9750$\pm$250 & 1.9$\pm$0.4 & 1.5$\pm$0.5 & 258$\pm$17 & 56$\pm$1\\
\hline
\end{tabular}
\begin{minipage}{85mm}

  \textbf{Notes.} The distance, effective temperature, mass and radius for HD 179218 and HD 141569 are from \citet{Montesinos09} and \citet{Fairlamb15}, respectively; the projected rotational velocity from \citet{Guimaraes06} and \citet{Mora01}, and the inclination to the line of sight from  \citet{Dent05} and \citet{Mayozer16}. 
\end{minipage} 
\end{table} 

HD 179218 is a group I object according with the \citet{Meeus01} classification, with the near-IR ($K$ band) already in excess and a stable, single-peaked H$\alpha$ profile \citep{Mendi11a,Mendi12}. The mass accretion rate is $<$ 5 $\times$ 10$^{-8}$ M$_{\odot}$ yr$^{-1}$ \citep{Mendi11b}. A possible companion is located at $\sim$ 2.5 $arcsec$ \citep[][i.e. larger than the field of view covered from the interferometric observations]{Thomas07,Wheelgright10}. \citet{Fedele08} showed that HD 179218 has two rings of dust at 1 and 20 $au$, and a compact gas emitting region between 1 and 6 $au$. A longitudinal magnetic field of 51 $\pm$ 30 $G$ was measured by \citet{Hubrig09}.  

HD 141569 is a Meeus group II star with an IR excess starting at wavelengths close to 10 $\mu$m and a stable, double-peaked H$\alpha$ emission line with the red peak slightly less intense than the blue \citep{Mendi11a,Mendi12}. The accretion rate is between $\sim$ 10$^{-7}$ and 10$^{-8}$ M$_{\odot}$ yr$^{-1}$ \citep{Mendi11b,Fairlamb15}. The object has two low-mass stellar companions at wide distances \citep[angular separations of 7.6 and 9 $arcsec$]{Weinberger99}. The transitional nature of HD 141569 has been verified from direct imaging, showing a huge hole depleted of dust extending up to $\sim$ 100 $au$ \citep{Weinberger99}. The external disc shows two ring-like belts at $\sim$ 210 and 380 $au$ and other structures \citep[see][and references therein]{Mayozer16} that could be indicative of the presence of giant planets \citep{Wyatt05,Reche09}. An additional inner ring ($<$ 100 $au$), spatially coincident with warm gas, was recently reported by \citet{Konishi16}. A very weak or non-existent longitudinal magnetic field was reported by \citet{Wade07} from several metallic lines. 

\section{Observations and data reduction}
\label{Sect:Observations}
Observations of HD 179218 and HD 141569 were taken at the CHARA array with the VEGA spectrometer on July 2015. The wavelength range covered was 6300-6700 \AA{}, at a spectral resolution $\sim$ 6000. The configurations used were S1S2, E1E2, and S2W2, with corresponding baselines of 34, 66, and 178 $m$. Calibration stars of similar spectral types and brightness (HD 181383 and HD 140075) were observed with the same observing configuration, using a CAL-SCI-CAL-SCI sequence with SCI exposure times of $\sim$ 25 min. Data reduction was carried out using the standard VEGA procedure \citep{Mourard09}. The first processing is done using a large spectral band of 30 $nm$ on which the spectral density of the short exposures are summed and the raw squared visibility is extracted. In all cases we found that the target is unresolved in this large spectral band. The second processing concerns the computation of the differential visibility by computing the cross-spectrum between the same large band and a narrow band of 2 $nm$ (equivalent spectral resolution of 320) moving across the large band. The differential visibility and differential phase are thus extracted and they help in putting constraints on the size and astrometric position of each small spectral band with respect to the large band. Instrumental artefacts were corrected for by dividing the observables of the target stars by those from the calibrators.    

\begin{figure}
\centering
 \includegraphics[width=8.4cm,clip=true]{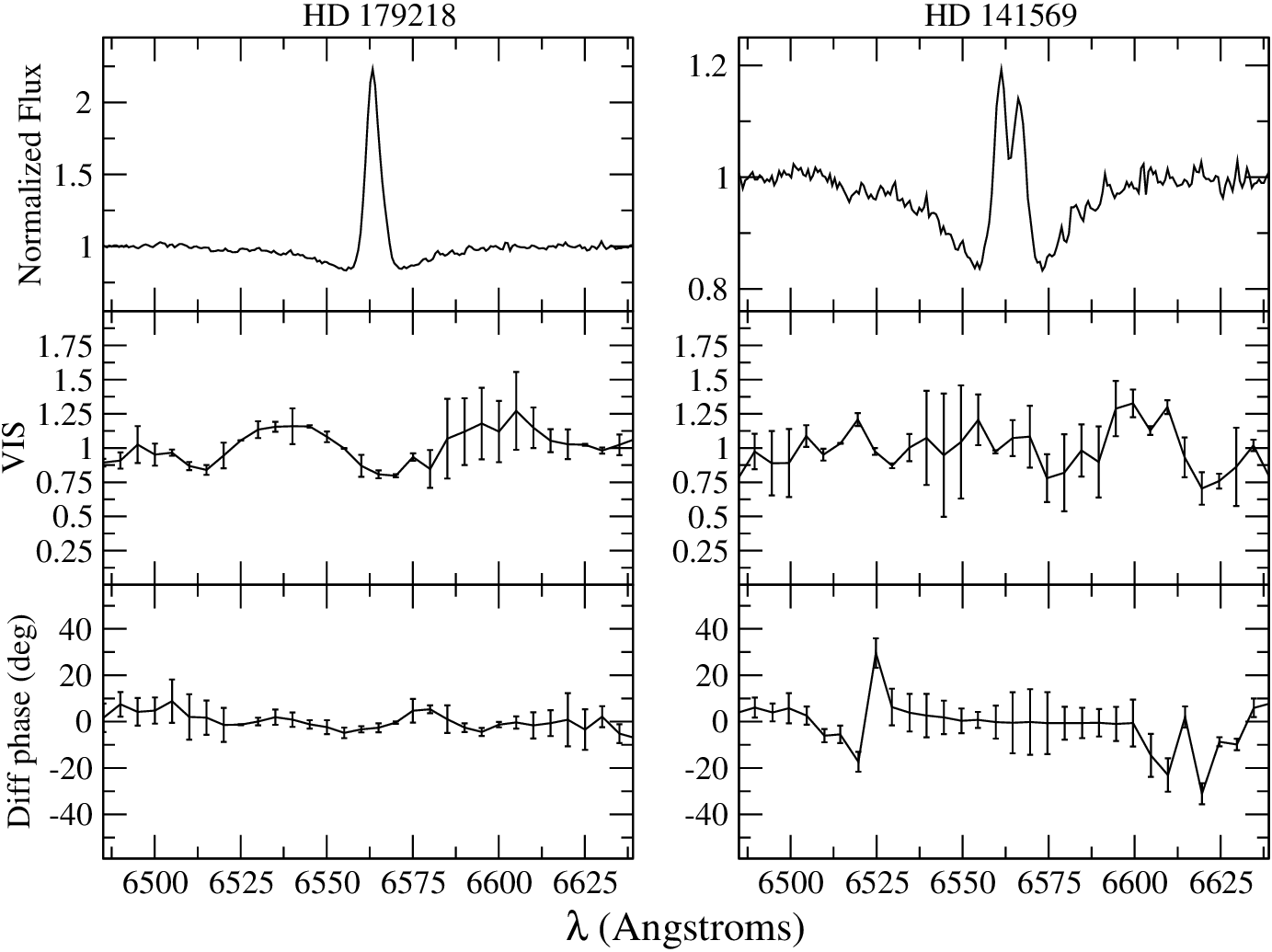}
\caption{CHARA normalized H$\alpha$ fluxes (note the different scales), visibilities, and differential phases for HD 179218 and HD 141569 taken with the 66 $m$ E1E2 baseline.}
\label{Figure:chara_results}
\end{figure}

Data taken with the S1S2 and S2W2 baselines were finally rejected because of the low signal to noise ratio resulting from poor observing conditions on 06 July 2015 and 12 July 2015. The E1E2 observations were taken on 07 July 2015 and are of good quality. Figure \ref{Figure:chara_results} shows the averaged results for the E1E2 baseline in terms of normalized H$\alpha$ line fluxes, visibilities, and differential phases. Error bars for the visibilities and differential phases come from the standard deviation of the SCI frames. The on-sky position angles (from north to east) are 297-305 $\degr$ and 300-320 $\degr$ for HD 179218 and HD 141569, respectively, the ranges coming from the spread in the $uv$ coverage. The H$\alpha$ profiles shown by HD 179218 and HD 141569 are single and double-peaked respectively, the observed emission with an equivalent width (EW) of -8.47 $\pm$ 0.05 and -3.2 $\pm$ 0.2 \AA{}. The H$\alpha$ line remains practically unaltered from previous observations \citep[e.g.][]{Mendi11a}. The visibilities randomly oscillate around 1 ($\pm$ $\sim$ 0.25), without showing any significant difference between the continuum and the H$\alpha$ line. This indicates that both the central stars and the line emitting regions are spatially unresolved. Similarly, the differential phases are randomly distributed around 0$\degr$ (with errors up to $\sim$ $\pm$ 20 $\degr$), showing no signs of asymmetries in the plane of the sky.

\section{Analysis}
\label{Sect:Analysis}
The fact that the H$\alpha$ line is unresolved sets an initial constraint on the size of the emitting region. The smallest angular distance that can be resolved from our observations is given by $\lambda$/2B $\sim$ 1 $mas$, with $\lambda$ = 6562.8 \AA{}, and B = 66 $m$ (baseline E1E2). At the distance of the target stars, this angular resolution indicates that the extent of the H$\alpha$ emitting region should be smaller than 0.21 $au$ (15R$_*$) for HD 179218, and $\sim$ 0.12 $au$ (16R$_*$) for HD 141569. That extent would reduce to only $\sim$ 6R$_*$ for both stars if the visibilities obtained with the S2W2 baseline are considered, although this value should be taken with caution given the poor quality of the S2W2 data. These results contrast with the two HAeBes that have been previously resolved with CHARA. The H$\alpha$ line of AB Aur was resolved even with a smaller baseline (34 $m$), indicating a extent of the line emitting region of at least $\sim$ 2 $mas$, or $\sim$ 0.3 $au$ at 144 $pc$ \citep{Rousselet10}. Regarding the binary MWC 361, \citet{Benisty13} estimated that the size of the H$\alpha$ emitting region is in between $\sim$ 0.2 and $\sim$ 0.6 $au$, depending on the position of the companion. Non-spherical or magnetically driven disc winds with a spatial extent larger than the disc truncation radius were invoked in those works to fit the observed properties of AB Aur and MWC 361. Two different physical scenarios that could explain the comparatively small H$\alpha$ emitting regions of HD 179218 and HD 141569 are discussed in the following. 

\subsection{Blackbody disc}
\label{Sect:disc}   

An additional constraint for the size of the H$\alpha$ emitting region can be derived assuming that this is an optically thick line characterized by a single, black-body temperature T$_{H\alpha}$. The line luminosity can then be estimated from the integration of the Planck's law, and can be expressed as L$_{H\alpha}$ = $\epsilon$A$_{H\alpha}$ $\sigma$ T$_{H\alpha}^4$, with A$_{H\alpha}$ the projected area of the line emitting region, $\sigma$ the Stefan-Boltzmann constant, and 0 $\leq$ $\epsilon$ $\leq$ 1 a correction factor that takes into account that the integration is carried out across the full width of the line. The value for $\epsilon$ can be derived numerically since
\begin{equation}
 \label{Eq:epsilon}
\epsilon = \frac{15}{\pi^4}\int_{u_2}^{u_1} \frac{u^3du}{e^u-1}; u_i =\frac{hc}{\lambda_i k T_{H\alpha}},
\end{equation}
with $\lambda_1$ and  $\lambda_2$ the shortest and longest wavelengths where the H$\alpha$ emission extends, and $h$, $c$ and $k$ the Planck's constant, the speed of light, and the Boltzmann's constant, respectively. Since the stellar luminosity is L$_{*}$ = A$_{*}$ $\sigma$ T$_{*}^4$, we have that L$_{H\alpha}$/L$_{*}$ = $\epsilon$(A$_{H\alpha}$/A$_{*}$)(T$_{H\alpha}$/T$_{*}$)$^4$. The line to stellar luminosity ratio can also be expressed as L$_{H\alpha}$/L$_{*}$ = (d/R$_*$)$^2$(F$_0$EW/$\sigma$T$_*^4$)10$^{-0.4m}$, with EW the line equivalent width, m the $R$-magnitude characterizing the continuum adjacent to the H$\alpha$ line and F$_0$ the corresponding zero-magnitude flux \citep[taken from][]{Besell79}. From the two previous expressions for L$_{H\alpha}$/L$_{*}$, the ratio between the area of the line emitting region and that of the star can be derived as A$_{H\alpha}$/A$_{*}$ = (F$_0$EW10$^{-0.4m}$/$\sigma$$\epsilon$T$_{H\alpha}^4$)(d/R$_*$)$^2$. Finally, the radial extent of the H$\alpha$ emitting region depends on the geometry assumed. If H$\alpha$ is emitted in a inclined disc extending from the stellar surface (assumed spherical) up to a stelleocentric radius r$_{H\alpha}$, then this is given by
\begin{equation}
 \label{Eq:size}
r_{H\alpha} = R_* \sqrt{\frac{F_0 EW 10^{-0.4m}}{\sigma \epsilon T_{H\alpha}^4 \cos i}\frac{d^2}{R_*^2} +1}.
\end{equation}
Other possible geometries for the line emitting region -e.g. spherical- provide smaller values of r$_{H\alpha}$.

Despite the simplicity of the assumptions, Eq. \ref{Eq:size} is similar to the radius of the inner gaseous discs that has been succesfully used to explain the hydrogen emission of supergiants, classical Be stars and, more recently, of early type HAeBe stars \citep[see e.g.][and references therein]{Grundstrom06,deWit08,Patel16}. A more realistic model would include a range of temperatures and optical depths, although the assumption of a single blackbody temperature should be enough to roughly describe an inner disc emitting an optically thick line such as H$\alpha$. Indeed, a dependence of the previous expression is the temperature characterizing the line emitting region, T$_{H\alpha}$. In principle, this is a free parameter that should lie within several thousand $K$ for hydrogen to be at least partially ionized. For T$_{H\alpha}$ = T$_*$, the total projected lengths of the H$\alpha$ emitting region ($\sim$ 2$r_{H\alpha}$) should be $\sim$ 2.5R$_*$ for both stars. This is below the spatial resolution limit of the baselines considered in this work, and therefore consistent with our observations. Higher temperatures would decrease r$_{H\alpha}$, whereas only very low, unlikely T$_{H\alpha}$ temperatures ($<$ 4000 $K$) would allow to spatially resolve the H$\alpha$ line emitting region. As a consequence, we conclude that a compact gaseous disc emitting like a blackbody at a temperature of several thousand $K$ could in principle be responsible of the H$\alpha$ emission for both HD 179218 and HD 141569. In addition, a double-peaked line profile like the one shown by HD 141569 is characteristic of keplerian-rotating discs. The peak emissions are located at around $\pm$ 115 km s$^{-1}$. This velocity corresponds to a Keplerian distance as close as $\sim$ 5R$_*$, where uncertainties from the spectral resolution and the error bars for the stellar masses and radii in Table \ref{Table:prop} have also been taken into account. A homogeneous disc would produce a double-peaked line profile with similar peak intensities, thus the fact that the red peak is less intense than the blue could in principle be explained from differeces in density between both sides of the rotating disc. However, the H$\alpha$ profile of HD 141569 is roughly stable on timescales ranging from several days to almost two decades \citep[see][]{Mendi11a}, which challenges the Keplerian disc as the only explanation for the H$\alpha$ emission. A relatively constant infall of material would cause a redshifted absorption which could explain why the red peak is always less intense than the blue. The inverse P-Cygni feature superimposed on some of the double-peaked H$\alpha$ profiles shown in \citet{Mendi11a} (not apparent in our CHARA observations) is consistent with this explanation.  Given that the inclination estimated for HD 179218 indicates that this system is closer to pole-on (Sect. \ref{Sect:MA}), a possible Keplerian disc could not be reflected by a characteristic double-peaked emission as the one shown by HD 141569, which is closer to edge-on.       

The previous procedure was repeated for the two other HAeBes resolved by CHARA in order to test whether their H$\alpha$ emission could also be consistent with the blackbody disc model. Eq. \ref{Eq:size} indicates that the corresponding projected size for AB Aur would be $\sim$ 0.04 $au$ (3.2R$_*$), i.e. an order of magnitude smaller than inferred from CHARA observations \citep{Rousselet10}. Apart from the distance and inclination indicated in this paper, we have assumed the stellar temperature (=T$_{H\alpha}$) and radius in \citet{DonehewBrit11} and the H$\alpha$ parameters in \citet{Costigan14}. Once again, only for unrealistically low T$_{H\alpha}$ temperatures ($<$ 3000 $K$) the blackbody disc could have been spatially resolved. Regarding MWC 361, the corresponding projected disc size would be $\sim$ 0.08 $au$ (2.4R$_*$), which is also an order of magnitude smaller than obtained from CHARA \citep{Benisty13}. The data necessary to apply Eq. \ref{Eq:size} were taken from that paper and from \citet{Hernandez04}. It is noted that in this case H$\alpha$ temperatures up to 6000 $K$ could make the blackbody disc large enough to be resolved. In conclusion, given the small emitting size compared to the one derived from CHARA, the blackbody disc hypothesis is not consistent with the observed properties of AB Aur, in contrast to our target stars. In the case of MWC 361, a ``cold'' blackbody disc with a temperature up to $\sim$ 6000 $K$ could also explain that H$\alpha$ is spatially resolved in this object. 

\subsection{Magnetospheric accretion}
\label{Sect:MA} 
 
A major constraint of the magnetospheric accretion (MA) scenario is that the H$\alpha$ emission line is generated in the magnetosphere, in magnetically driven flows at a radial distance not further than the co-rotation radius \citep{Shu94}. For the stellar parameters, projected rotational velocities and inclinations assumed here (Sect. \ref{Sect:Sample}), the co-rotation radii are 2.39 R$_*$ and 1.36 R$_*$ for HD 179218 and HD 141569, well below the radial extent that can be resolved with CHARA. There are only two previous examples in the literature for which MA has been used to reproduce the H$\alpha$ emission shown by HAeBes \citep[UX Ori and BF Ori in][respectively]{Muzerolle04,Mendi11b}. HI line profiles in the near-IR of the HAeBe stars VV Ser and HD 58647 were also succesfully reproduced from combined MA-wind models by \citet{GarciaLopez16} and \citet{Kurosawa16}, respectively \citep[see also][]{Tambo14,Tambo16}. The remaining lines of evidence supporting that MA could work at least for several HAeBes mainly come from MA shock modelling of the near-UV excess \citep{Muzerolle04,Mendi11b,Fairlamb15}, spectro-polarimetry \citep{Vink02,Mottram07} and statistical analysis of line profiles \citep{Cauley14,Cauley15}.   

In order to stablish whether MA is able to reproduce the H$\alpha$ profiles of HD 179218 and HD 141569, we have applied the line modelling described in \citet{Muzerolle01}. Specific details can be consulted there \citep[see also][]{Muzerolle04,Mendi11b}, but a general description is included in the following. The model calculates the atomic level populations of free-falling gas channeled along magnetic field lines in a dipolar, axisymmetric field emerging from the star. Input data are the stellar parameters (M$_*$, R$_*$, T$_*$), the mass accretion rate ($\dot{M}_{\rm acc}$), the maximum temperature of the gas (T$_{H\alpha}$$^{max}$), the inclination ($i$), and the size of the magnetosphere in terms of the minimum and maximum radial distances from which the gas is channelled \citep[r$_{mi}$ and r$_{ma}$; see Fig. 1 in][]{Muzerolle01}. The stellar parameters and inclinations used for the modelling are those in Table \ref{Table:prop}, the mass accretion rates were varied within the limits provided by \citet{Mendi11b} and \citet{Fairlamb15} from MA shock modelling (Sect. \ref{Sect:Sample}), r$_{ma}$ was fixed to the co-rotation radius, and r$_{mi}$ was varied as a free parameter. Regarding T$_{H\alpha}$$^{max}$, the actual physics involved in heating the gas is still not understood, so this remains the least well-constrained of all the parameters. As a first-order approach it has been assumed that T$_{H\alpha}$$^{max}$ is similar within a few hundred $K$ to the MA shock temperatures provided in \citet{Mendi11b} and \citet{Fairlamb15} for HD 179218 and HD 141569, which is similar to the gas temperatures in the top of the range found for TTs \citep{Muzerolle01}. We refer the reader to \citet{Muzerolle01} and \citet{Muzerolle04}, where the influence of the different input parameters on the shape of the modelled line profiles were extensively discussed.  

\begin{figure}
\centering
 \includegraphics[width=8.4cm,clip=true]{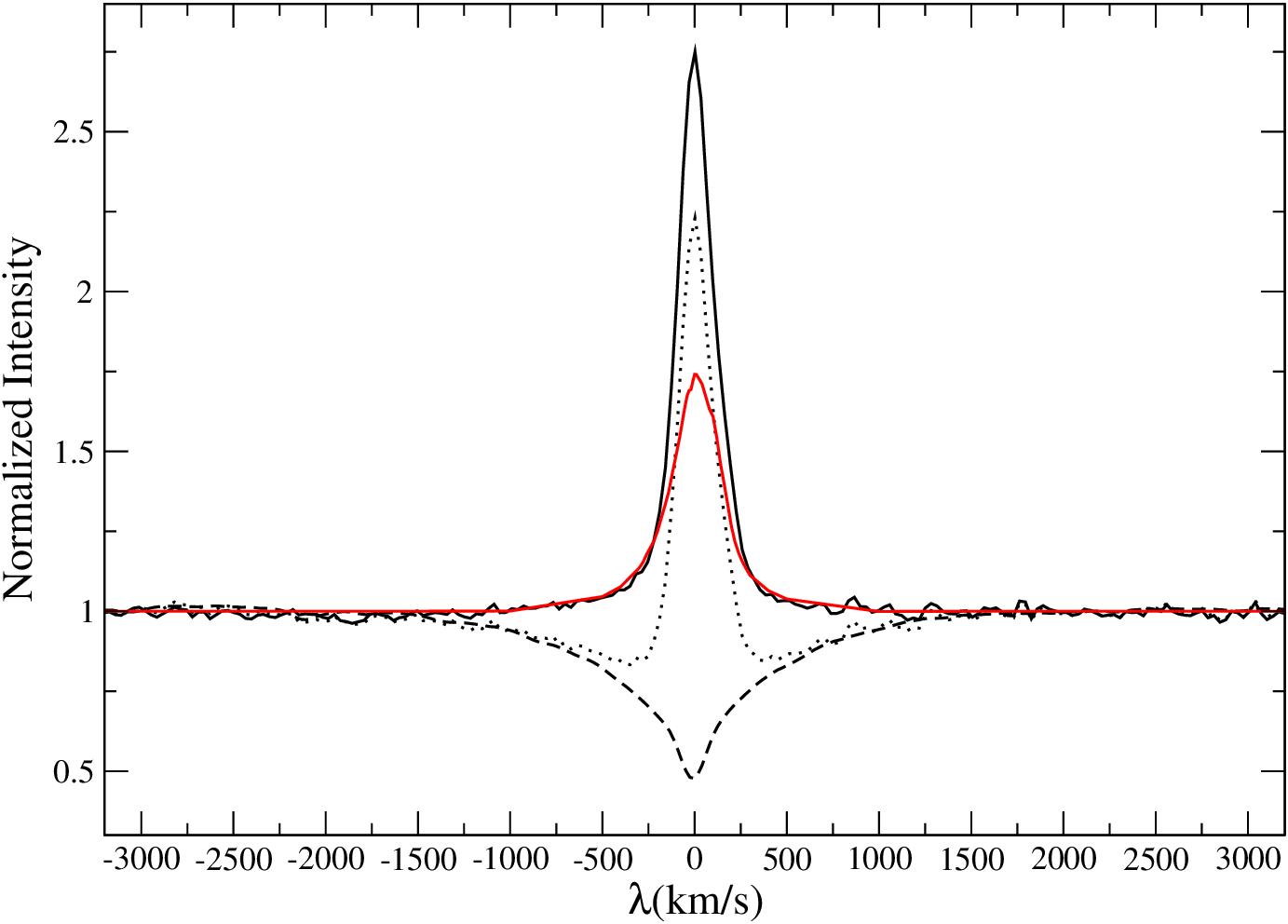}
\caption{The observed H$\alpha$ spectrum HD 179218 is indicated with the dotted line; the photospheric and the non-photospheric contributions with dashed and solid lines, respectively. The result from MA modelling is indicated with the red line (see text).}
\label{Figure:model_obs}
\end{figure}

\begin{figure}
\centering
 \includegraphics[width=8.4cm,clip=true]{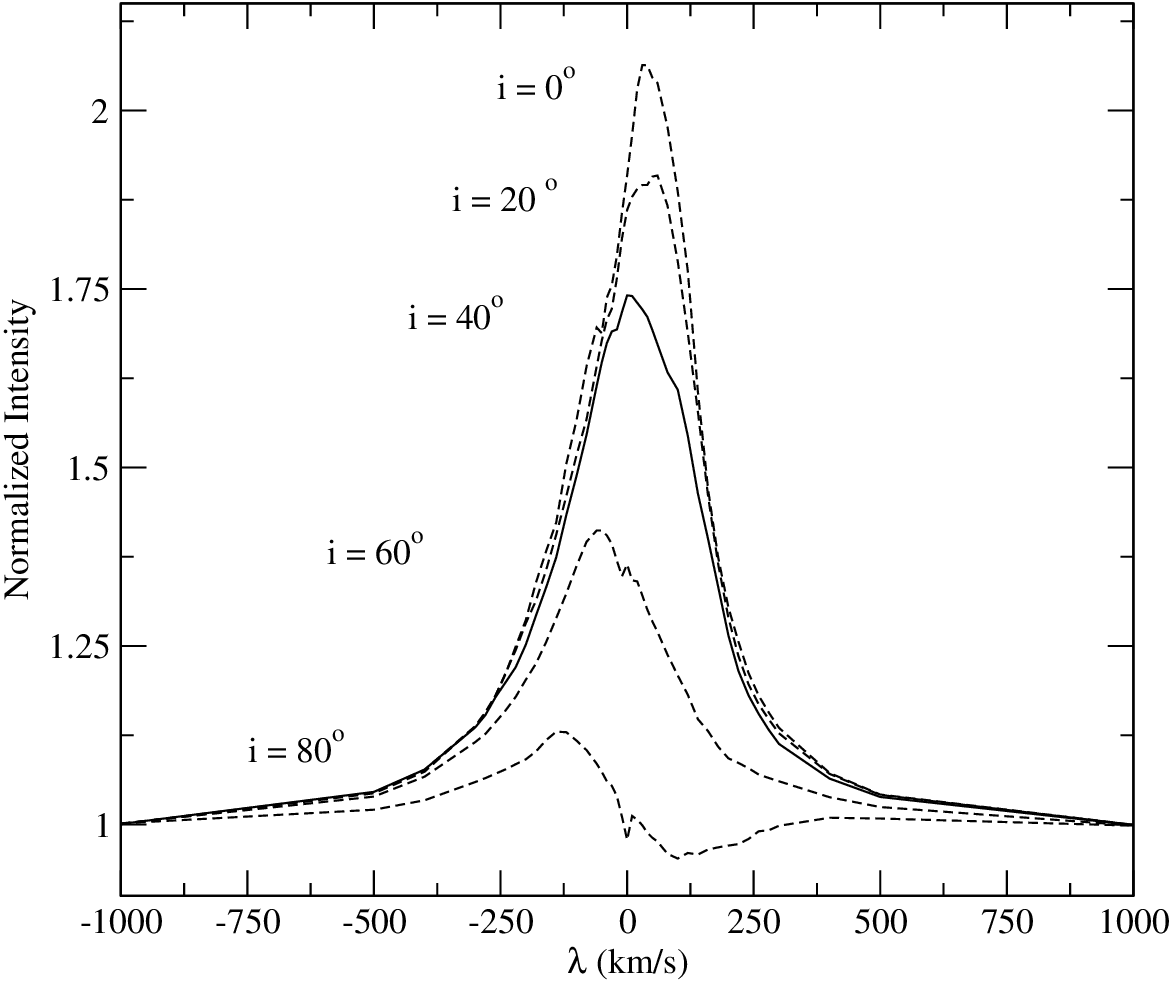}
\caption{The previous modelled line profile for HD 179218 (solid line), considering additional inclinations to the line of sight (dashed lines), as indicated.}
\label{Figure:model_incl}
\end{figure}

Fig. \ref{Figure:model_obs} shows the results of this modelling for HD 179218. The H$\alpha$ profile observed with CHARA is plotted with a dotted line and includes the contribution of the photospheric absorption. This has been removed using a Kurucz synthetic profile corresponding to a star with the same effective temperature, surface gravity, and rotational velocity (dashed line). The resulting, non-photospheric H$\alpha$ line is indicated with a black solid line. The best fit model is plotted with a red, solid line. The corresponding model parameters are the ones in Table \ref{Table:prop}, and $\dot{M}_{\rm acc}$ = 1.7 $\times$ 10$^{-8}$ M$_{\odot}$ yr$^{-1}$,  T$_{H\alpha}$$^{max}$ = 13000 $K$, r$_{mi}$ = 2.35R$_{*}$, r$_{ma}$ = 2.40R$_{*}$. The modelled peak emission is smaller than observed (although larger fluxes result from inclinations smaller than 40$\degr$, see below). The radiative transfer methodology (using the Sobolev approximation) may break down in the densest, lowest-velocity part of the flow, which may affect the poor fit near the line center. However, MA modelling is not only able to reproduce at least $\sim$ 60 $\%$ of the non-photospheric H$\alpha$ flux of HD 179218, but also fits the observed line wings. This indicates that MA is able to explain the bulk of the H$\alpha$ emission in this object, and specially the high-velocity component. It is noted that the H$\alpha$ broadening is strongly influenced by the Stark effect, for which models alternative to MA may also be able to reproduce the line wings as far as they include hot enough emitting regions. Remarkably enough, the current MA line modelling is consistent with the previous MA shock modelling of this star \citep{Mendi11b}, both providing the same values for $\dot{M}_{\rm acc}$ and T$_{H\alpha}$$^{max}$. Moreover, MA line modelling confirms the inclination derived for this object by \citet{Dent05}. Fig. \ref{Figure:model_incl} shows that despite low inclinations ($i$ $<$ 30$\degr$) produce larger fluxes more similar than observed, the resulting peak emission is redshifted. On the other hand, H$\alpha$ models with high inclinations ($>$ 50$\degr$) are blueshifted and do not fit the red wing of the line, even showing inverse P-Cygni profiles for edge-on configurations (i$>$ 70$\degr$).    

Regarding HD 141569, not only the corresponding range of input parameters mentioned above are unable to reproduce the observed H$\alpha$ profile, but apparently no set of input parameters can do it. Variations in the mass accretion rates of several orders of magnitude, changes in T$_{H\alpha}$$^{max}$ of several thousand $K$, and a wide range of possible r$_{mi}$ values have been attempted without success. As discussed in \citet{Muzerolle04} \citep[see also][]{Mendi11b}, very high stellar rotation rates as the one shown by HD 141569 (Table \ref{Table:prop}) are problematic for the MA model used. Those result in very small co-rotation radii, for which the emission volume associated to such a small accretion flow is not enough to match the observed H$\alpha$ emission regardless of any other combination of parameters. As argued by \citet{Muzerolle04}, geometries deviating from a standard dipole may reproduce the observations. Therefore, further theoretical development on how MA line modelling could deal with high rotational velocities (typically, vsini $>$ 150 km s$^{-1}$) is necessary.   

In spite of these results, we confirm that MA cannot be the primary origin of H$\alpha$ emission in the two other HAeBes previously observed with CHARA. If the bulk of the line emission in AB Aur and MWC 361 were mainly generated in the magnetosphere, the line emitting region would even be smaller than in the previous, blackbody disc case, for which it would have been impossible to spatially resolve the line as it was done by \citet{Rousselet10} and \citet{Benisty13}. Moreover, a P-Cygni H$\alpha$ profile such as the one shown by AB Aur is clearly inconsistent with accretion. As argued by \citet{Rousselet10}, a full spherical wind providing matter along the line of sight would produce too much blue-shifted absorption with respect to the observed P-Cygni feature, leaving the X-disk or disk-wind geometry as the best suited scenario. 

\section{Conclusions}
\label{Sect:Conclusions}
Our H$\alpha$ interferometric observations of the HAeBe stars HD 179218 and HD 141569 adds to the two previously reported (AB Aur and MWC 361), revealing that different phenomena could be originating the emission in this type of objects. The line can be emitted in a very compact region of a few stellar radii from hot blackbody discs (HD 179218 and HD 141569). The bulk of the H$\alpha$ emission can be explained from magnetospheric accretion in the case of HD 179218, but the current model cannot deal with high rotational velocities as the one shown by HD 141569. A more extended emission can be explained from non-spherical winds (AB Aur and MWC 361), but a cold blackbody disc could also cause the H$\alpha$ line of MWC 361. Despite the small sample of HAeBes that can be observed with current optical spectro-interferometric instrumentation, the variety reported matches with previous observations of the Br$\gamma$ transition over a wider sample, where depending on the star the emission can also be compact, extended, associated to discs, accretion or winds.

\section*{Acknowledgements}

The authors thank the anonymous referee for his/her useful comments on the original manuscript, which helped us to improve the
paper. CHARA Array time was granted through the NOAO community-access program (NOAO PropID: 2015A-0098; PI: I. Mendigut\'\i{}a). The CHARA Array is funded by the National Science Foundation through NSF grant AST-1211129 and by Georgia State University through the College of Arts and Sciences. 












\bsp	
\label{lastpage}
\end{document}